\documentclass[prd,showpacs,preprintnumbers,twocolumn,amsmath,amssymb,nofootinbib,floatfix]{revtex4}

\usepackage{graphicx}
\usepackage{epsfig}
\usepackage{epstopdf}
\usepackage{color,xcolor}
\usepackage{bm,booktabs,dcolumn}
\usepackage{longtable,lscape}
\usepackage{txfonts}
\usepackage{overpic}
\usepackage{float}
\usepackage{array}
\usepackage{multirow}
\usepackage{subfigure}
\usepackage{microtype}
\usepackage{pdfpages}
\usepackage{tikz}

\usepackage{amsmath,amssymb,amsfonts,amsbsy,mathrsfs,dsfont}
\usepackage{cases}
\usepackage{slashed}
\usepackage{feynmf}

\usepackage[normalem]{ulem}

\makeatletter
\@addtoreset{equation}{section}
\makeatother

\newcommand{\nc}{\newcommand}
\nc{\tj}[1]{\textcolor{red}{Tianjin: #1}}

\usepackage{soul}

\allowdisplaybreaks

\usepackage[colorlinks, citecolor=blue,anchorcolor=red,menucolor=red, linkcolor=red,filecolor=red,runcolor=red,urlcolor=blue,frenchlinks=true]{hyperref}

\begin{document}
\title{$3^{++}$ light meson family}
\author{Hao Chen$^{1,5}$}\email{chenhao$\_$qhnu@outlook.com}
\author{ Yan-Yue Fan $^{1}$}
\author{Yun-Hai Zhang$^{2}$}\email{wlxzyh@163.com}
\author{Cheng-Qun Pang$^{3,4}$}\email{xuehua45@163.com}
\affiliation{
$^1$ College of Physics and Electronic Information Engineering, Qinghai Normal University, Xining 810000, China\\
$^2$ College of Physics and Electronic Engineering, Heze University, Heze 274015, China\\
$^3$ School of Physics and Optoelectronic Engineering, Ludong University, Yantai 264000, China\\
$^4$Lanzhou Center for Theoretical Physics, Key Laboratory of Quantum Theory and Applications of MoE, and Key Laboratory of Theoretical Physics of Gansu Province, Lanzhou University, Lanzhou, Gansu 730000, China \\
$^5$ Academy of Plateau Science and Sustainability, Xining 810016, China
}

\date{\today}

\begin{abstract}
As members of the $J^{PC}=3^{++}$ light meson family, the assignments of the $a_3(1875)$, $a_3(2030)$, $a_3(2275)$, $f_3(2050)$, and $f_3(2300)$ states remain unclear. In this work, we investigate the mass spectra and the Okubo-Zweig-Iizuka-allowed two-body strong decays of the $3^{++}$ light meson family using the modified Godfrey-Isgur quark model and the quark-pair creation model. We explore the identification of the $a_3(1875)$, $a_3(2030)$, $a_3(2275)$, $f_3(2050)$, and $f_3(2300)$ states. Additionally, we predict the masses and widths of the higher excitations in the $3^{++}$ light meson family.

\end{abstract}
\maketitle

\section{Introduction}\label{sec1}

The spectroscopy of light mesons plays a crucial role in exploring the nonperturbative aspects of Quantum Chromodynamics. Among the various meson families, higher-spin states provide essential tests of quark model predictions. In this context, investigations of the  $a_3$ and $f_3$ meson families are of particular interest.

The $a_3$ and $f_3$ meson {families are} characterized by {the quantum numbers} $J^{PC}=3^{++}$ (P and C denote spatial parity and charge conjugation parity, respectively), which corresponds to spin $S=1$ and total angular momentum $J=3$.
In particular, the isovector states with \(I^G=1^-\), denoted as the \(a_3\) family, and the isoscalar states with \(I^G=0^+\), denoted as the \(f_3\) family, have attracted sustained attention. Several members, such as \(a_3(2030)\), \(a_3(2275)\), \(f_3(2050)\), and \(f_3(2300)\), are listed in the Particle Data Group (PDG), with their masses, widths, and branching fractions given by a variety of experiments ~\cite{ParticleDataGroup:2024cfk}.

The $3^{++}$ light meson family traces its origins to about 25 years ago.
An early partial-wave analysis (PWA) by the E852 Collaboration of in the reaction $\pi^{-} p \!\to\! \pi^{+}\pi^{-}\pi^{-} p$ at 18 GeV/$c$ reported an $a_3(1875)$ resonance~\cite{Chung:2002pu}. Subsequently, the Crystal Barrel Collaboration at LEAR reported two higher isovector states in $\bar{p}p$ annihilation: $a_3(2030)$ and $a_3(2275)$~\cite{Anisovich:2001pn,Anisovich:2001pp}.

For the isoscalar sector, a pronounced $3^{++}$ signal identified as $f_3(2050)$ was observed in $\bar pp$ annihilation into several final states, notably $\bar pp\!\to\!\eta\pi^{0}\pi^{0}$, $\pi^{0}\pi^{0}$, $\pi^{+}\pi^{-}$, $\eta\eta$, and $\eta\eta'$~\cite{Anisovich:2000ut}. A heavier isoscalar partner, $f_3(2300)$, was subsequently introduced via a dedicated PWA of $\bar pp\!\to\!\Lambda\bar\Lambda$ data~\cite{Bugg:2004rj}.

\par

Regarding spectroscopic assignments, $a_3(2030)$ and $a_3(2275)$ line up naturally on the $^3\!F_3$ Regge trajectory and are commonly interpreted as $1^3\!F_3$ and $2^3\!F_3$  states~\cite{Masjuan:2012gc, Pang:2015eha}, while the relativistic quark model spectra by Ebert \textit{et al.} assign $a_3(1875)$ as the $a_3$ ground state and identify $a_3(2030)$ and $a_3(2275)$ as the first and the second radial excitations, respectively~\cite{Ebert:2009ub}. Moreover, Refs.~\cite{Masjuan:2012gc, Pang:2015eha} suggest that $a_3(1875)$ and $a_3(2030)$ may be the same state, serving as the ground state of the $a_3$ family, with $a_3(2275)$ as its first radial excitation.

Refs.~\cite{Anisovich:2000ut,Afonin:2007aa,Pang:2015eha} interpret $f_3(2050)$ as the $1^3\!F_3$ ground state and $f_3(2300)$ as its first radial excitation, whereas Ref.~\cite{Masjuan:2012gc} proposes that $f_3(2050)$ and $f_3(2300)$ correspond to the first and second radial excitations, respectively, thereby implying an unobserved ground $f_3$ {state} near $1.7$ GeV. A quasipotential relativistic quark model calculation by Ebert \textit{et al.} further suggests that $f_3(2300)$ could be the ground state of the $f_3$ family with a dominant $s\bar{s}$ component~\cite{Ebert:2009ub}.

For the isoscalar members of {the} $3^{++}$ family,  $f_3$ and $f_3^\prime$ are mixtures of the nonstrange and strange flavor basis states,
\begin{equation}\label{mixingns}
\begin{pmatrix}
f_3(nF)\\[2pt]
f_3'(nF)
\end{pmatrix}
=
\begin{pmatrix}
\cos\theta_{n\,^{3}\!F_{3}} & \sin\theta_{n\,^{3}\!F_{3}} \\
-\sin\theta_{n\,^{3}\!F_{3}} & \cos\theta_{n\,^{3}\!F_{3}}
\end{pmatrix}
\begin{pmatrix}
|n\bar n\rangle \\
|s\bar s\rangle
\end{pmatrix},
\end{equation}
with $n\bar{n}=\frac{u\bar{u}+(-1)^Id\bar{d}}{\sqrt{2}}$, $u$, $d$, and $s$ denote the ``up'', ``down'', and ``strange'' quarks,  respectively.

Ref.~\cite{Dudek:2011tt}
reports a nonstrange--strange mixing angle for the $f_3(1F)$ mesons of
$\theta_{1^3F_3}=3\pm2^\circ$. Testing  this  mixing angle of $f_3(1F)$ mesons and studying the mixing angle of other $f_3$ mesons from decay properties can help us to understand their internal structure.

As previously mentioned, the classification of the $3^{++}$ light mesons remains controversial, making the establishment of them an interesting and important research issue.

Over the past 40 years, the Godfrey-Isgur (GI) potential model has been widely used to investigate meson spectra~\cite{Godfrey:1985xj,Godfrey:2016nwn}. More recently,  the modified GI (MGI) model, which incorporates color screening effects~\cite{Laermann:1986pu,Born:1989iv,Knechtli:2000df,Chao:1992et,Ding:1993uy}, has been successfully applied to  doubly heavy flavor mesons, light mesons and baryons~\cite{Song:2015nia,Wang:2018rjg,Wang:2019mhs,Weng:2024roa}.
For describing the nature of the higher spin $3^{++}$ light family, the MGI model is adopted  to study their spectral properties.

In our previous work, we adopted the MGI  model to predict the spectrum properties of {the} $5^{++}$ and  $4^{++}$  meson families~\cite{Pang:2019ttv,Pang:2025umu}.
In parallel, we adopted the quark-pair creation (QPC) model, also known as the $^3P_0$ model~\cite{Micu:1968mk,Ackleh:1996yt}, to predict their two-body strong decays. In this work, we adopted these two models to study the spectra and two-body {strong} decays of the $3^{++}$ family. We hope that our effort will be helpful in revealing the internal structure of $a_3(1875)$, $a_3(2030)$, $a_3(2275)$, $f_3(2050)$, and $f_3(2300)$, and in establishing the $3^{++}$ meson family.

\par
The paper is organized as follows: In Sec. \ref{sec2}, we briefly review the MGI model and the $^3P_0$ model. Sec. \ref{sec3} presents our numerical results for the $3^{++}$ family: we first verify the assignment of the   $a_3(1875)$, $a_3(2030)$, $a_3(2275)$, $f_3(2050)$, and $f_3(2300)$, and then provide predictions for higher $3^{++}$ states. Conclusions are summarized in Sec.~\ref{sec4}.

\section{Models employed in this work} \label{sec2}

In the following part, spectroscopy (masses and wave functions) is obtained from a screened and relativistic quark potential (the MGI model), and the Okubo-Zweig-Iizuka (OZI)-allowed two-body decay widths are computed in the QPC model. These two models are introduced below.

\vspace{0.5em}

\subsection{The modified GI model}

Building on the GI model, which achieved notable success in describing relativistic meson spectra--especially the low-lying states~\cite{Godfrey:1985xj},  Song \textit{et al.} proposed the MGI model~\cite{Song:2015nia}. By incorporating a screened confining potential, the MGI model provides an improved description of excited charmed and charmed-strange mesons in Ref.~\cite{Song:2015nia}. Since its introduction, it has been applied to light meson spectroscopy~\cite{Pang:2018gcn,Wang:2022xxi,Wang:2021abg,Feng:2022hwq,Wang:2024yvo,Wang:2024lba}, double-heavy quarkonia~\cite{Wang:2018rjg,Wang:2019mhs}, and baryons~\cite{Weng:2024roa}. The Hamiltonian of the MGI model reads:
\begin{equation}\label{hh}
\tilde{H}=\sum_i\sqrt{m_i^{\,2}+\mathbf{p}^{\,2}}+\tilde{V}^{\mathrm{eff}},
\end{equation}
where $m_i$ denotes the constituent mass of the quark (or antiquark) and $\mathbf{p}$ is the relative {three-}momentum.
Throughout, we take $m_{u(d)}=0.162~\mathrm{GeV}$ for ``up'' (``down'') quark and $m_s=0.377~\mathrm{GeV}$ for ``strange'' quark (see Table~\ref{MGI}) .
The effective potential is decomposed as
\begin{equation}\label{V}
\tilde{V}^{\mathrm{eff}}
=\tilde{G}_{12}
+\tilde{V}^{\mathrm{cont}}
+\tilde{V}^{\mathrm{tens}}
+\tilde{V}^{\mathrm{so(v)}}
+\tilde{S}_{12}(r)
+\tilde{V}^{\mathrm{so(s)}},
\end{equation}
with the following interpretation of each term:
\begin{itemize}
  \item $\tilde{G}_{12}$: the (smeared) one-gluon-exchange (OGE) Coulomb interaction;
  \item $\tilde{V}^{\mathrm{cont}}$: the short-range spin-spin contact piece;
  \item $\tilde{V}^{\mathrm{tens}}$: the tensor interaction from OGE;
  \item $\tilde{V}^{\mathrm{so(v)}}$: the vector spin-orbit contribution;
  \item $\tilde{S}_{12}(r)$: the screened confining potential that encodes color screening;
  \item $\tilde{V}^{\mathrm{so(s)}}$: the scalar spin-orbit contribution.
\end{itemize}
Explicit expressions and detailed discussions of these potential terms are presented below.

In the nonrelativistic limit, the spin-independent central pieces read
\begin{equation}
\tilde{G}(r)=-\sum_{k=1}^3\frac{4\alpha_k}{3r}\left[\frac{2}{\sqrt{\pi}}\int_{0}^{\gamma_k r} e^{-x^2}\,dx\right],
\end{equation}
with the effective strong coupling coefficients $(\alpha_1,\alpha_2,\alpha_3)=(0.25,\,0.15,\,0.20)$ and the range parameters $(\gamma_1,\gamma_2,\gamma_3)=\bigl(1/2\,\mathrm{GeV},\,\sqrt{10}/{2}\,\mathrm{GeV},\,\sqrt{1000}/{2}\,\mathrm{GeV}\bigr)$~\cite{Godfrey:1985xj}, and
\begin{equation}
S(r)=\frac{b\,(1-e^{-\mu r})}{\mu}+c,
\end{equation}
where we adopt the screening parameter $\mu=0.0779~\mathrm{GeV}$ from our previous work~\cite{Wang:2024lba}.
This parameter controls the strength of color screening--absent in the original GI model and thus constitutes a key refinement in the MGI framework.
We set $b=0.222~\mathrm{GeV}^2$ for the string tension and $c=-0.228~\mathrm{GeV}$ for the constant term~\cite{Wang:2024lba}.

The relativistic effects are incorporated in two ways. First, the constituents' internal motion and the nonlocality of the $q\bar q$ interaction are modeled through a mass-dependent Gaussian smearing function $\rho_{ij}(\mathbf r-\mathbf r')$:
\begin{equation}
\label{smearing}
\rho_{ij}(\mathbf r-\mathbf r')=\frac{\sigma_{ij}^3}{\pi^{3/2}}
\exp\!\big[-\sigma_{ij}^2(\mathbf r-\mathbf r')^2\big],
\end{equation}
with
\begin{align}
\label{smearing_sigma}
\sigma_{ij}^2
=\sigma_0^2\!\left[\tfrac12+\tfrac12\!\left(\frac{4m_i m_j}{(m_i+m_j)^2}\right)^{\!4}\right]
+s^2\!\left(\frac{2m_i m_j}{m_i+m_j}\right)^{\!2},
\end{align}
where $\sigma_0=1.791~\mathrm{GeV}$ and $s=0.711$ are the universal parameters, with values taken from Ref.~\cite{Wang:2024lba}.
This smearing is applied to the central OGE Coulomb kernel $G(r)$ and the screened confinement $S(r)$ as:
\begin{equation}
\begin{split}
\tilde{G}_{ij}(r)=&\int {\rm{d}}^3{\bf r}^\prime \rho_{ij}({\bf r}-{\bf r}^\prime)G(r^\prime)
=-\sum\limits_{k=1}^3\frac{4\alpha_k }{3r}{\rm erf}(\tau_{kij}r),
\end{split}
\end{equation}
where
\begin{equation}
\tau_{kij}=\frac{1}{\sqrt{\frac{1}{\sigma_{ij}^2}+\frac{1}{\gamma_k^2}}},
\end{equation}
and
\begin{eqnarray}
\tilde{S}_{12}(r)&=& \int {\rm{d}}^3 {\bf r}^\prime
\rho_{12} ({\bf r}-{\bf r}^\prime)S(r^\prime)\nonumber\\
&=& \frac{b}{\mu r}\Bigg[r+e^{\frac{\mu^2}{4 \sigma^2}+\mu r}\frac{\mu+2r\sigma^2}{2\sigma^2}\Bigg(\frac{1}{\sqrt{\pi}}
\int_0^{\frac{\mu+2r\sigma^2}{2\sigma}}e^{-x^2}{\rm{d}}x-\frac{1}{2}\Bigg) \nonumber\\
&&-e^{\frac{\mu^2}{4 \sigma^2}-\mu r}\frac{\mu-2r\sigma^2}{2\sigma^2}\Bigg(\frac{1}{\sqrt{\pi}}
\int_0^{\frac{\mu-2r\sigma^2}{2\sigma}}e^{-x^2}{\rm{d}}x-\frac{1}{2}\Bigg)\Bigg]  \nonumber \\
&&+c, \nonumber\label{Eq:pot}
\end{eqnarray}
where $\sigma\equiv\sigma_{12}$.
The spin-dependent interactions (contact, tensor, spin-orbit) are then constructed from these smeared central kernels (see below).

Second, relativistic kinematics is incorporated through momentum-dependent similarity factors that reduce to unity in the nonrelativistic limit.
Accordingly, the smeared Coulomb term is modified as
\begin{equation}
\tilde G_{12}(r)\ \longrightarrow\
\left(1+\frac{\mathbf p^2}{E_1E_2}\right)^{\!1/2}\!
\tilde G_{12}(r)\!
\left(1+\frac{\mathbf p^2}{E_1E_2}\right)^{\!1/2},
\end{equation}
where $E_{1(2)}=\sqrt{m_{1(2)}^2+\mathbf p^2}$. In parallel, the spin-dependent operators built from the smeared kernels are modified by semirelativistic corrections via
\begin{equation}
\label{vsoij}
\tilde V^{\,i}_{\alpha\beta}
=
\left(\frac{m_\alpha m_\beta}{E_\alpha E_\beta}\right)^{\!1/2+\epsilon_i}
\tilde V^{\,i}_{\alpha\beta}(r)
\left(\frac{m_\alpha m_\beta}{E_\alpha E_\beta}\right)^{\!1/2+\epsilon_i},
\end{equation}
where $i$ labels the four spin-dependent components:
$i=\{\mathrm{cont},\,\mathrm{tens},\,\mathrm{so(v)},\,\mathrm{so(s)}\}$,
and the corresponding exponents
$\epsilon_i=\{\epsilon_c,\,\epsilon_t,\,\epsilon_{\rm so(v)},\,\epsilon_{\rm so(s)}\}$  represent the relativistic corrections to $\tilde{V}^{\mathrm{cont}}$, $\tilde{V}^{\mathrm{tens}}$, $\tilde{V}^{\mathrm{so(v)}}$, and $\tilde{V}^{\mathrm{so(s)}}$, respectively, and with values taken from Ref.~\cite{Wang:2021abg}.

Then the explicit spin-dependent potentials expressed in terms of the smeared OGE and confinement kernels read
\begin{equation}\label{Vcont}
\tilde V^{\mathrm{cont}}
=\frac{2\,\mathbf S_1\!\cdot\!\mathbf S_2}{3m_1m_2}\,\nabla^2\tilde G^{\,c}_{12},
\end{equation}
\begin{equation}\label{Vtens}
\tilde V^{\mathrm{tens}}
=-\frac{3(\mathbf S_1\!\cdot\!\hat{\mathbf r})(\mathbf S_2\!\cdot\!\hat{\mathbf r})-\mathbf S_1\!\cdot\!\mathbf S_2}{3m_1m_2}
\left(\frac{d^2}{dr^2}-\frac{1}{r}\frac{d}{dr}\right)\tilde G^{\,t}_{12},
\end{equation}
\begin{equation}\label{Vsov}
\tilde V^{\mathrm{so(v)}}=
\frac{\mathbf S_1\!\cdot\!\mathbf L}{2m_1^2}\frac{1}{r}\frac{d\tilde G^{\rm so(v)}_{11}}{dr}
+\frac{\mathbf S_2\!\cdot\!\mathbf L}{2m_2^2}\frac{1}{r}\frac{d\tilde G^{\rm so(v)}_{22}}{dr}
+\frac{(\mathbf S_1+\mathbf S_2)\!\cdot\!\mathbf L}{m_1m_2}\frac{1}{r}\frac{d\tilde G^{\rm so(v)}_{12}}{dr},
\end{equation}
\begin{equation}\label{Vsos}
\tilde V^{\mathrm{so(s)}}=
-\frac{\mathbf S_1\!\cdot\!\mathbf L}{2m_1^2}\frac{1}{r}\frac{d\tilde S^{\rm so(s)}_{11}}{dr}
-\frac{\mathbf S_2\!\cdot\!\mathbf L}{2m_2^2}\frac{1}{r}\frac{d\tilde S^{\rm so(s)}_{22}}{dr}.
\end{equation}

\begin{table}[htbp]
\renewcommand{\arraystretch}{1.5}
\caption{Parameters of the MGI model~\cite{Wang:2024lba}. $m_{u(d)}$ and $m_s$ are the masses of the $u(d)$ and $s$ quarks (or antiquarks), $\mu$ is the screening parameter, $b$ is the confining  parameter, and $c$ is the vacuum constant. $\sigma_0$ is the universal parameter in Eq.~(\ref{smearing_sigma}), while $s$ is a parameter related to heavy quarkonium masses. $\epsilon_c$, $\epsilon_t$, $\epsilon_{\rm so(v)}$, and $\epsilon_{\rm so(s)}$ represent the relativistic corrections to the potential terms $\tilde{V}^{\mathrm{cont}}$, $\tilde{V}^{\mathrm{tens}}$, $\tilde{V}^{\mathrm{so(v)}}$, and $\tilde{V}^{\mathrm{so(s)}}$, respectively. $\gamma=5.1$ is fitted with the total width of
$a_3(1F)$, $a_3(2F)$, $f_3(1F)(\theta_{1^3F_3}=0)$, and $f_3(2F)(\theta_{2^3F_3}=0)$.
\label{MGI}}
\begin{center}
\begin{tabular}{cccc}
\hline\hline
Parameter &  value &Parameter &  value  \\
 \midrule[0.7pt]
$m_{u(d)}$(GeV)    &0.162    &{$s$ }          &{0.711}\\
$m_s$ (GeV)       &0.377    &$\mu$ (GeV)          &0.0779 \\
$b$ (GeV$^2$)     &0.222    &$c$ (GeV)            &$-0.228$\\
$\epsilon_c$      &-0.137   &$\epsilon_{so(v)}$     &0.0550\\
$\epsilon_{so(s)}$  &0.366    &$\epsilon_t$         &0.493\\
 {$\sigma_0$ (GeV)}   &{1.791}  &$\gamma$   &5.1\\
  \hline\hline
\end{tabular}
\end{center}
\end{table}

We obtain the light meson spectrum by expanding the states in a simple-harmonic-oscillator (SHO) basis.
The SHO eigenfunctions are
\begin{equation}
\begin{split}
\psi_{nLM_L}^{\mathrm{SHO}}(\mathbf r)=R_{nL}^{\mathrm{SHO}}(r,\beta)\,Y_{LM_L}(\Omega_r),\\
\psi_{nLM_L}^{\mathrm{SHO}}(\mathbf p)=R_{nL}^{\mathrm{SHO}}(p,\beta)\,Y_{LM_L}(\Omega_p),
\end{split}
\end{equation}
with radial parts
\begin{align}\label{1.3}
R_{nL}^{\mathrm{SHO}}(r,\beta)
&=N_{nL}\,\beta^{3/2}(\beta r)^{L}\,e^{-\,\beta^2 r^2/2}\,
L_{n-1}^{L+1/2}(\beta^2 r^2),\\
R_{nL}^{\mathrm{SHO}}(p,\beta)
&=\frac{(-1)^{n-1}(-i)^L}{\beta^{3/2}}\,N_{nL}\,
e^{-\,p^2/(2\beta^2)}\left(\frac{p}{\beta}\right)^{L}
L_{n-1}^{L+1/2}\!\left(\frac{p^2}{\beta^2}\right),
\end{align}
and normalization
\begin{align}\label{1.4}
N_{nL}=\sqrt{\frac{2\,(n-1)!}{\Gamma(n+L+1/2)}}.
\end{align}
Here $Y_{LM_L}(\Omega)$ are spherical harmonics, $L_{n-1}^{L+1/2}(x)$ are associated Laguerre polynomials, and $\Gamma(\cdot)$ denotes the gamma function.

In this representation, the radial wave functions of the light $q\bar{q}$
mesons under study are expanded as
\begin{align}\label{expand}
R_{nL}(r)&=\sum_{k=1}^{N_{\max}} C_k\,R_{kL}^{\mathrm{SHO}}(r,\beta),\\
R_{nL}(p)&=\sum_{k=1}^{N_{\max}} C_k\,R_{kL}^{\mathrm{SHO}}(p,\beta),
\end{align}
where the coefficients $C_k$ are obtained by diagonalizing the Hamiltonian in Eq.~(\ref{hh}), and we retain $N_{\max}=21$ basis functions in our numerical calculations.
The SHO basis depends on a single scale parameter $\beta$; for each meson $i$ we fix $\beta_i$ variationally by minimizing the eigenvalue $E_{nL}$, i.e.,
$\partial E_{nL}/\partial \beta_i=0$ and $\partial^2 E_{nL}/\partial \beta_i^2>0$.
Part of related $\beta$ values can be found in Ref.~\cite{Pang:2025esm}.
The resulting spatial wave functions from the MGI model are then used as inputs for the strong-decay analysis.

\subsection{The QPC model}\label{QPC}

The QPC model was first proposed by Micu~\cite{Micu:1968mk} and later developed by the Orsay group~\cite{LeYaouanc:1972ae,LeYaouanc:1973xz,LeYaouanc:1974mr,LeYaouanc:1977gm,LeYaouanc:1977ux}.
It has since been widely applied to the study of OZI-allowed two-body strong decays of mesons~\cite{vanBeveren:1982qb,Titov:1995si,Ackleh:1996yt,Blundell:1996as,Bonnaz:2001aj,Zhou:2004mw,Lu:2006ry,Zhang:2006yj,Luo:2009wu,Sun:2009tg,Liu:2009fe,Sun:2010pg,Rijken:2010zza,Ye:2012gu,Wang:2012wa,He:2013ttg,Sun:2013qca,Pang:2018gcn,Wang:2022juf,Wang:2022xxi,Li:2022khh,Li:2022bre,Wang:2020due,Pang:2017dlw,Wang:2024yvo,Wang:2021abg,feng:2021igh,Feng:2022hwq}.
Within this framework, the transition operator $\mathcal T$ creates a quark-antiquark pair (labeled 3 and 4) from the vacuum with quantum numbers $^{2S+1}L_J={}^{3}P_0$. For the process $A \rightarrow B+C$, $\mathcal T$ can be expressed as

{\begin{align}\label{gamma}
\mathcal{T} = & -3\gamma \sum_{m}\langle 1m;1~-m|00\rangle\int d \mathbf{p}_3d\mathbf{p}_4\delta ^3 (\mathbf{p}_3+\mathbf{p}_4) \nonumber \\
 & ~\times \mathcal{Y}_{1m}\left(\frac{\mathbf{p}_3-\mathbf{p}_4}{2}\right)\chi _{1,-m}^{34}\phi _{0}^{34}
\left(\omega_{0}^{34}\right)_{ij}b_{3i}^{\dag}(\mathbf{p}_3)d_{4j}^{\dag}(\mathbf{p}_4).
\end{align}
Here, the parameter $\gamma$ in the QPC model characterizes the strength of $q\bar q$-pair creation from the vacuum.
In the present work, the value $\gamma = 5.1$ is determined by fitting to the total widths of
$a_3(1F)$, $a_3(2F)$, $f_3(1F)$ $(\theta_{1^3F_3}=0)$, and $f_3(2F)$ $(\theta_{2^3F_3}=0)$.
The function $\mathcal{Y}_l^m(\mathbf p)\equiv p^lY_l^m(\theta_p,\phi_p)$ denotes a solid harmonic,
while $\chi$, $\phi$, and $\omega$ represent the spin, flavor, and color wave functions, respectively.
$\mathbf p_3$ and $\mathbf p_4$ are the three-momenta of the quark and antiquark created from the vacuum,
and $i$ and $j$ denote their color indices.
With this transition operator, the decay amplitudes of mesons can be systematically evaluated within the QPC framework.
The amplitude $\mathcal{M}^{M_{J_{\mathrm A}}M_{J_{\mathrm B}}M_{J_{\mathrm C}}}$ is defined as

\begin{equation}
\langle {\mathrm{BC}}|\mathcal{T}|{\mathrm{A}} \rangle = \delta ^3({\mathbf{P}_{\mathrm{B}}+\mathbf{P}_{\mathrm{C}})}\mathcal{M}^{{M}_{J_{\mathrm{A}}}M_{J_{\mathrm{B}}}M_{J_{\mathrm{C}}}},
\end{equation}
where $\mathbf{P}_{\mathrm B}$ and $\mathbf{P}_{\mathrm C}$ are the three-momenta of mesons $\mathrm B$ and $\mathrm C$ in the rest frame of the initial meson $\mathrm A$, and $M_{J_i}$ ($i=\mathrm A,\mathrm B,\mathrm C$) denotes the corresponding magnetic quantum numbers.
The partial width takes the form
\begin{eqnarray}
\Gamma_{{\mathrm{A}}\to {\mathrm{BC}}}&=&\frac{\pi}{4} \frac{|\mathbf{P}|}{m_{\mathrm{A}}^2}\sum_{J,L}|\mathcal{M}^{JL}(\mathbf{P})|^2,
\end{eqnarray}
with $m_{\mathrm A}$ the mass of $\mathrm A$, $\mathbf{P}\equiv \mathbf{P}_{\mathrm B}$ , $L$ the relative orbital angular momentum between $\mathrm B$ and $\mathrm C$, and $J$ the total angular momentum obtained by coupling $\mathbf{J}=\mathbf{J}_{\mathrm B}+\mathbf{J}_{\mathrm C}$.
The partial-wave amplitude $\mathcal{M}^{JL}(\mathbf{P})$ is related to the  amplitude $\mathcal{M}^{{M}_{J_{\mathrm{A}}}M_{J_{\mathrm{B}}}M_{J_{\mathrm{C}}}}$ via the Jacob-Wick transformation~\cite{Jacob:1959at}:

\begin{equation}
\begin{aligned}
\mathcal{M}^{JL}(\mathbf{P}) = &\frac{\sqrt{4\pi(2L+1)}}{2J_{\mathrm{A}}+1}\sum_{M_{J_{\mathrm{B}}}M_{J_{\mathrm{C}}}}\langle L0;JM_{J_{\mathrm{A}}}|J_{\mathrm{A}}M_{J_{\mathrm{A}}}\rangle \\
    &\times \langle J_{\mathrm{B}}M_{J_{\mathrm{B}}};J_{\mathrm{C}}M_{J_{\mathrm{C}}}|{J_{\mathrm{A}}}M_{J_{\mathrm{A}}}\rangle \mathcal{M}^{M_{J_{{\mathrm{A}}}}M_{J_{\mathrm{B}}}M_{J_{\mathrm{C}}}},
    \end{aligned}	
\end{equation}
in which
\begin{equation}
\begin{aligned}
 &\mathcal{M}^{M_{J_{\mathrm{A}}} M_{J_{\mathrm{B}}} M_{J_{\mathrm{C}}}}\\
 &=\gamma \sum_{\substack{M_{L_{\mathrm{A}}}, M_{S_{\mathrm{A}}}, M_{L_{\mathrm{B}}},\\ M_{S_{\mathrm{B}}}
M_{L_{\mathrm{C}}, M_S}, m}}\left\langle L_{\mathrm{A}} M_{L_{\mathrm{A}}} S_{\mathrm{A}} M_{S_{\mathrm{A}}} \mid J_{\mathrm{A}} M_{J_{\mathrm{A}}}\right\rangle \\
& \times\left\langle L_{\mathrm{B}} M_{L_{\mathrm{B}}} S_{\mathrm{B}} M_{S_{\mathrm{B}}} \mid J_{\mathrm{B}} M_{J_{\mathrm{B}}}\right\rangle\left\langle L_{\mathrm{C}} M_{L_{\mathrm{C}}} S_{\mathrm{C}} M_{S_{\mathrm{C}}} \mid J_{\mathrm{C}} M_{J_{\mathrm{C}}}\right\rangle \\
& \times\langle 1 m 1-m \mid 00\rangle\left\langle\chi_{S_{\mathrm{B}} M_{S_{\mathrm{B}}}}^{14} \chi_{S_{\mathrm{C}} M_{S_{\mathrm{C}}}}^{32} \mid \chi_{S_{\mathrm{A}} M_{S_{\mathrm{A}}}}^{12} \chi_{1-m}^{34}\right\rangle \\
& \times\left[\left\langle\phi_{\mathrm{B}}^{14} \phi_{\mathrm{C}}^{32} \mid \phi_{\mathrm{A}}^{12} \phi_0^{34}\right\rangle I\left(\mathbf{P}, m_1, m_2, m_3\right)\right.+ \\
& \left.(-1)^{1+S_{\mathrm{A}}+S_{\mathrm{B}}+S_{\mathrm{C}}+L_{\mathrm{C}}}\left\langle\phi_{\mathrm{B}}^{32} \phi_{\mathrm{C}}^{14} \mid \phi_{\mathrm{A}}^{12} \phi_0^{34}\right\rangle I\left(-\mathbf{P}, m_2, m_1, m_3\right)\right], \\
\end{aligned}
\end{equation}
with {the overlap integral}
\begin{equation}
\begin{aligned}
&I\left(\mathbf{P}, m_1, m_2, m_3\right)\\
&=\int d^3 \mathbf{k} \psi_{\mathrm{B}}^*(\mathbf{k}+U \mathbf{P}) \psi_{\mathrm{C}}^*(\mathbf{k}+V\mathbf{P} ) \psi_{\mathrm{A}}(\mathbf{k}-\mathbf{P}) \mathcal{Y}_1^m(\mathbf{k}),
\end{aligned}
\end{equation}
{where}
\begin{equation*}
\begin{aligned}
U=\frac{m_3}{m_1+m_3}, V=\frac{m_3}{m_2+m_3}{.}
\end{aligned}
\end{equation*}
Here $m_1$ and $m_2$ are the constituent quark and antiquark masses in the initial meson $\mathrm A$. In our calculations we take
$(m_1,m_2)=(0.377,0.162)\,\mathrm{GeV}$ for $s\bar n$ (or $n\bar s$), $(0.162,0.162)\,\mathrm{GeV}$ for $n\bar n$, and $(0.377,0.377)\,\mathrm{GeV}$ for $s\bar s$. The mass of the created quark (or antiquark) is $m_3=0.162\,\mathrm{GeV}$ for $n\bar n$ creation and $m_3=0.377\,\mathrm{GeV}$ for $s\bar s$.

The meson spatial wave functions entering the above overlap integrals in the strong-decay calculations are obtained from the MGI model and employed throughout our QPC analysis.

For the final states, the mixing between strange mesons with natural parity ($L=J$) is described by
\begin{equation}\label{anglek1}
\left( \begin{matrix}
	|K(nL)\rangle \\
	|K^\prime( nL)\rangle \\
\end{matrix}\right) =
\left( \begin{matrix}
	\textrm{$\cos\theta_{nL}$} & \textrm{$\sin\theta_{nL}$} \\
	\textrm{$-\sin\theta_{nL}$} & \textrm{$\cos\theta_{nL}$} \\
\end{matrix}\right)
\left( \begin{matrix}
	|K(n^1L_L)\rangle \\
	|K(n^3L_L)\rangle \\
\end{matrix}\right),
\end{equation}
where $\theta_{nL}$ denotes the mixing angle between the $K(n^{1}L_L)$ and $K(n^{3}L_L)$ configurations. The mixing angle for the $P$-wave case is taken as $\theta_{1P}=-34^\circ$~\cite{Cheng:2013cwa}, while for higher orbital excitations it is determined by $\theta_{nL} = -\arctan\!\left(\sqrt{\frac{L}{L+1}}\right)$~\cite{Asghar:2019qjl}.

The flavor wave functions of isoscalar mesons are expressed in the mixing form
\begin{equation}\label{mixingns-quark-flavor ba}
\left( \begin{matrix}
	X \\
	X^\prime \\
\end{matrix}\right) =
\left( \begin{matrix}
	\textrm{$\cos\theta_{x}$} & \textrm{$\sin\theta_{x}$} \\
	\textrm{$-\sin\theta_{x}$} & \textrm{$\cos\theta_{x}$} \\
\end{matrix}\right)
\left( \begin{matrix}
	|n\bar{n}\rangle \\
	|s\bar{s}\rangle \\
\end{matrix}\right),
\end{equation}
where $X$ and $X^\prime$ denote the two isoscalar partners (e.g., $\eta$ and $\eta^\prime$), $\theta_x$ is the mixing angle in the quark-flavor basis.
The flavor-mixing parameters of the isoscalar mesons used in this work are adopted from Ref.~\cite{Pang:2025esm}.

\begin{table*}[ht]
\renewcommand{\arraystretch}{1.5}
\centering
\caption{Mass spectrum of $3^{++}$ states. The units of width and mass are MeV, and ``Expt. Inf.'' represents experimental information. \label{mass}}
\[\begin{array}{ccccccccccccccc}
\hline
\hline
&  & I^G n^{2S+1}L_J (flavor)& \text{State}  & \text{This work} & \text{GI~\cite{Steph:1985ff}} & \text{Ebert~\cite{Ebert:2009ub}} & \text{Expt. Inf.~\cite{ParticleDataGroup:2024cfk}}
\\\midrule[0.7pt]
&  & 1^- 1^3F_3(n\bar n) &a_3(2030)    &1955  &2046  &1910 & M=1874\pm43\pm96, \text{$\Gamma$=}385\pm121\pm114\text{~\cite{Chung:2002pu}}   \\
&&&&&&&M=2031\pm12,\text{$\Gamma$=}150\pm18\text{~\cite{Anisovich:2001pn}}     \\
&&&&&&&B(a_3 (1875) \to f_2\pi)/B(a_3 (1875) \to \rho\pi)=0.8\pm0.2\text{~\cite{Chung:2002pu}}\\
&&&&&&&B(a_3 (1875) \to \rho_3(1690)\pi)/B(a_3 (1875) \to \rho\pi)=0.9\pm0.3\text{~\cite{Chung:2002pu}}     \\
&  &  1^- 2^3F_3(n\bar n) &a_3(2275)    &2234&2444 & 2191   &  M=2275\pm 35,\text{$\Gamma$=}350_{-50}^{+100}\text{~\cite{Anisovich:2001pp}}\\
&  &  1^- 3^3F_3 (n\bar n) & a_3(3F)   &2454 &2800& ... &  ...                \\
&  & 1^-  4^3F_3 (n\bar n) & a_3(4F)     &2631  &3124 &...&... \\
\hline
&  & 0^+ 1^3F_3 (n\bar n) &f_3(2050)    &1955  &2046  &1910 & M=2048\pm8,\text{$\Gamma$=}213\pm34\text{~\cite{Anisovich:2000ut}}     \\
&  &  0^+ 2^3F_3(n\bar n)  &f_3(2300)    &2234&2444 & 2191   &  M=2334\pm 25,\text{$\Gamma$=}200\pm20\text{~\cite{Bugg:2004rj}}\\
&  &  0^+ 3^3F_3 (n\bar n) & f_3(3F)   &2454 &2800& ... &  ...                \\
&  &  0^+ 4^3F_3 (n\bar n) & f_3(4F)     &2631  &3124 &...&... \\
\hline
&  & 0^+ 1^3F_3 (s\bar s) &f_3^\prime(1F)    &2166  &2234  &2215 &    ... \\
&  &  0^+ 2^3F_3 (s\bar s) &f_3^\prime(2F)    &2452&2623 & 2585   &   ...\\
&  &  0^+ 3^3F_3 (s\bar s) &f_3^\prime(3F)   &2682 &2972& ... &  ...                \\
&  &  0^+ 4^3F_3 (s\bar s) & f_3^\prime(4F)     &2870  &3290 &...&... \\
 \hline
 \hline
\end{array}\]
\end{table*}

\section{Numerical results and phenomenological analysis}\label{sec3}
The mass spectrum of the $3^{++}$ light meson family is calculated with the MGI model using parameters in Table~\ref{MGI} and listed in Table~\ref{mass}.
The OZI-allowed two-body strong decays are summarized in Tables~\ref{a31F2F}-\ref{2H4F}. We now turn to a phenomenological analysis of the spectrum and the decay information of the $3^{++}$ light meson family.

\subsection{Mass spectra analysis}
Alongside predictions from the GI-model~\cite{Godfrey:1985xj}, Ebert \textit{et al.}~\cite{Ebert:2009ub}, and available experimental information, our results are listed in Table~\ref{mass}.

For the $1^3F_3(n\bar n)$ states, our predictions by the MGI model are $M[a_3(2030)]=1955$ MeV and $M[f_3(2050)]=1955$ MeV, being lower than the PDG values $2031\pm12$  and $2048\pm8$ MeV by $\sim80$ and $\sim90$ MeV, respectively~\cite{Anisovich:2001pn,Anisovich:2000ut}. Our result for $a_3$ is between the two experimental values of $1874\pm43\pm96$ MeV and $2031\pm12$ MeV~\cite{Chung:2002pu,Anisovich:2001pn}. In comparison, the GI model (2046 MeV) is consistent with the experimental values of $2031\pm12$  and $2048\pm8$ MeV~\cite{Anisovich:2001pn,Anisovich:2000ut}, while the relativistic quasipotential result of Ebert
\textit{et al.} (1910 MeV) agrees with the mass of $a_3 (1875)$~\cite{Chung:2002pu}.

Our results of the $2^{3}F_{3}(n\bar n)$ states are $M[a_{3}(2275)]=2234$ MeV and $M[f_{3}(2300)]=2234$ MeV (Table~\ref{mass}), which lie $\sim40$ and $\sim100$ MeV below the experimental values, $2275\pm35$ and $2334\pm25$ MeV, respectively. The GI values of 2444 MeV for both states are larger than the experimental values by $\sim170$ and $\sim110$ MeV, while those values by the Ebert \textit{et al.} quasipotential, 2191 MeV, are lower by $\sim80$ and $\sim140$ MeV, respectively. Thus for $a_{3}(2275)$ and $f_{3}(2300)$, our predictions lie between GI and Ebert \textit{et al.} and are closer to the experimental values.

The mainly $s\bar s$ isoscalar partners $f_3^\prime(1F)$ and $f_3^\prime(2F)$ have MGI predictions of 2166 and 2452 MeV, which are lower than the GI values of 2234 and 2623 MeV by 70 and 170 MeV, and below the Ebert \textit{et al.} results of 2215 and 2585 MeV by about 50 and 130 MeV, respectively.

For higher excitations, our results by the MGI model systematically lowers values from the GI model, consistent with screening effects.
As for $a_{3}(3F)$, our prediction is 2454 MeV, which lies 350 MeV below the GI value (2800 MeV).
The $a_{3}(4F)$ has a predicted mass of 2631 MeV, sitting 490 MeV beneath 3124 MeV for the GI model.
Compared with the GI mass at 2800 MeV, our result for $f_{3}(3F)$ is 2454 MeV, lower by 350 MeV.
Our calculation places $f_{3}(4F)$ at 2630 MeV, which is 490 MeV lower than the GI estimate of 3124 MeV.
The expected mass of 2682 MeV for $f_{3}^\prime(3F)$ is lower by 290 MeV than the GI result of 2972 MeV.
The $f_{3}^\prime(4F)$ mass is 2870 MeV, which is 420 MeV below the GI prediction of 3290 MeV.

{
In summary, the MGI model reproduces the observed $1F/2F$ levels with deviations of about 50-100 MeV from the experimental values and predicts substantially lower higher-$F$ excitations than the GI model, as expected once color screening is included.
} The mass spectra we obtained  support that $a_3(1875)$ and $a_3(2030)$ are the same state, serving as the ground state of the $a_3$ family, with $a_3(2275)$ as its first radial excitation, which is consistent with Ref.~\cite{Masjuan:2012gc} and our previous work~\cite{Pang:2015eha}. In addition, $f_3(2050)$ is the ground state of the $f_3$ family. And $f_3(2300)$ can be interpreted as the first radial excitation of the $f_3$. In the next section, we will further discuss these assignments with two-body strong decay calculations and provide channel-by-channel predictions for all the above states.

\subsection{Strong decay analysis}
We now turn to the investigation of the OZI-allowed two-body strong decays of the $3^{++}$ light meson family, whose results are presented in Tables~\ref{a31F2F}-\ref{2H4F}. The $q\bar q$-pair creation strength $\gamma=5.1$, introduced in Sec.~\ref{QPC}, is used consistently for all $a_3$, $f_3$, and $f_3^\prime$ states. In the decay tables, we emphasize the main modes and omit negligible contributions, adopting the criterion ${\rm Br}<0.5\%$ for suppression.

\subsubsection{Predictions of the ground and the first excited states of the light $3^{++}$ meson family}

As for $a_3(1F)$, the total width we obtained is about 220 MeV. The decay is dominated by $\pi\rho_3(1690)$, with a branching ratio of $41\%$ and a partial width of $91$ MeV. The next most important channels are $\pi\rho$ and $\pi f_2$, whose branching ratios are 11\% and 10\% and partial widths are 24 MeV and 23 MeV, respectively. The ratio $B(a_3 (1875) \to \pi f_2)/B(a_3 (1875) \to \pi\rho)\simeq0.9$ by our calculation agrees with the experimental value $0.8\pm0.2$ reported in Ref.~\cite{Chung:2002pu} within the experimental uncertainty.
Additionally, the ratio $B(a_3 (1875) \to \pi\rho_3(1690))/B(a_3 (1875) \to \pi\rho)$ in Ref.~\cite{Chung:2002pu} is about $0.9\pm0.3$, while our calculation gives the value of 3.7. See Table~\ref{a31F2F} for the remaining channels and details.

The total width of $a_3(2F)$ state we predicted is 160 MeV, which is approximately 140 MeV lower than the experimental value of $350_{-50}^{+100}$ MeV~\cite{Anisovich:2001pp}. The main decay modes are $\pi\rho_3(1990)$, $\pi\rho_3(1690)$, and $\pi f_2$, with branching ratios of 28\%, 12\%, and 8.0\%, and partial widths of 44, 18, and 13 MeV, respectively. All other channels are also shown in Table~\ref{a31F2F} for details.

\begin{table}[htbp]
\renewcommand{\arraystretch}{1.4}
\centering
\caption{Total and partial decay widths of $a_3(1F)$ and $a_3(2F)$. ``Br'' denotes the branching ratio of the decay channel. The unit of width is MeV. Channels with branching ratios below 0.5\% are omitted.
\label{a31F2F}}
\[\begin{array}{cccccc}
\hline
\hline
 \text{} & {a_3(1F)} & \text{} & \text{} & {a_3(2F)} & \text{} \\
 \midrule[0.7pt]
 \text{Channel} & \text{Br($\%$)} & \text{Width} & \text{Channel} & \text{Br($\%$)} & \text{Width} \\
 \hline
  \text{Total} & 100 & 222 & \text{Total} & 100 & 158 \\
 \pi \rho _3\text{(1690)} & 40.8 & 90.6 & \pi \rho _3\text{(1990)} & 27.9 & 44.2 \\
 \pi \rho  & 10.9 & 24.2 & \pi \rho _3\text{(1690)} & 11.6 & 18.4 \\
 \pi f_2 & 10.2 & 22.7 & \pi f_2 & 7.98 & 12.6 \\
 \rho h_1 & 8.57 & 19 & \text{$\pi \rho $(1450)} & 5.57 & 8.82 \\
 \rho \omega  & 7.74 & 17.2 & \pi \rho  & 4.68 & 7.41 \\
 \eta a_2 & 5.73 & 12.7 & \text{$\rho \pi $(1300)} & 3.77 & 5.96 \\
 \pi b_1 & 2.88 & 6.41 & \rho a_2 & 3.73 & 5.9 \\
 KK_2{}^* & 2.58 & 5.73 & \pi f_2\text{(1640)} & 3.59 & 5.69 \\
 \rho a_1 & 1.84 & 4.09 & \rho h_1 & 3.58 & 5.66 \\
 \pi f_1 & 1.62 & 3.6 & \rho a_1 & 2.76 & 4.36 \\
 \pi \eta _2 & 1.39 & 3.09 & \pi f_4\text{(2050)} & 2.73 & 4.32 \\
 \omega b_1 & 1.16 & 2.57 & \rho \omega  & 2.64 & 4.18 \\
 \text{$\pi \rho $(1450)} & 1.13 & 2.51 & \pi f_0\text{(1500)} & 2.46 & 3.9 \\
 KK^* & 0.864 & 1.92 & \pi f_1\text{(2P)} & 2.14 & 3.38 \\
 \text{...} & \text{...} & \text{...} & \pi \rho _2\text{(1D)} & 1.93 & 3.05 \\
 \text{...} & \text{...} & \text{...} & \eta a_2 & 1.86 & 2.94 \\
 \text{...} & \text{...} & \text{...} & \pi b_1\text{(2P)} & 1.83 & 2.89 \\
 \text{...} & \text{...} & \text{...} & \omega b_1 & 1.79 & 2.84 \\
 \text{...} & \text{...} & \text{...} & \text{$\pi \rho $(1900)} & 0.797 & 1.26 \\
 \text{...} & \text{...} & \text{...} & \text{$\pi \rho $(1700)} & 0.685 & 1.08 \\
 \text{...} & \text{...} & \text{...} & \text{$\rho \omega $(1420)} & 0.678 & 1.07 \\
 \text{...} & \text{...} & \text{...} & KK^*\text{(1410)} & 0.624 & 0.989 \\
 \text{...} & \text{...} & \text{...} & \pi \eta _2 & 0.576 & 0.913 \\
 \text{...} & \text{...} & \text{...} & \pi b_3\text{(2030)} & 0.544 & 0.861 \\
 \hline
\hline
 \end{array}\]
\end{table}

Fig.~\ref{f3(1F)} shows how the total decay widths of $f_3(1F)$ (dashed curve) and $f_3^\prime(1F)$ (solid curve) vary with the flavor-mixing angle $\theta$.
The blue band indicates the experimental range ($213\pm34$ MeV)~\cite{Anisovich:2000ut}, plotted alongside our $^3P_0$ calculation for direct comparison.
For $f_3(1F)$, the width shows a dome-like dependence on $\theta$, reaching values close to the experimental central line around $\theta\simeq 0$ and gradually decreasing as $|\theta|$ increases toward $90^\circ$.
The calculated width overlaps the SPEC band~\cite{Anisovich:2000ut} within the angular interval: $-34^\circ\lesssim\theta\lesssim42^\circ$. Our predicted total width at $\theta=0$ lies within the experimental range. Hence, $\theta=0$ is adopted as the representative mixing angle for subsequent analysis. Ref.~\cite{Dudek:2011tt} reports $\theta_{1^3F_3}=3\pm2^\circ$, numerically consistent with our choice and also sits within the same overlap region. Regarding $f_3^\prime(1F)$, the variation is opposite in phase: a pronounced minimum occurs near $\theta\simeq 0$ while the width increases as $|\theta|\to 90^\circ$.
{The total decay width of $f_3^\prime(1F)$ is between about 109 MeV and 187 MeV with $\theta$ ranging from $-34^\circ$ to $42^\circ$.}

Numerically, when we take $\theta=0$, $f_3(1F)$ has a total decay width of 230 MeV.
The main decay channels are $\pi a_2$, $\rho\rho$, and $\rho b_1$,
with branching ratios of $30\%$, $17\%$, and $11\%$, corresponding to partial widths of $70$, $39$, and $26$ MeV, respectively.
All other modes like $K K_2^{*}$ at 3.1\% and $K K^{*}$ below 1.0\% can be found in Table~\ref{1F2F}.

According to Table~\ref{1F2Fp}, we obtain the decay width of 110 MeV for $f_3^\prime(1F)$ at $\theta=0$, with $K K_2^{*}$ being the dominant channel contributing 46\% of the branching ratio and 51 MeV of the partial width.
$K K^{*}$ provides 23\% of the branching ratio with a partial width of 26 MeV, while $K^{*}K^{*}$ accounts for 15\% with 17 MeV.

The total decay widths of $f_3(2F)$ (dashed curve) and $f_3^\prime(2F)$ (solid curve) as functions of the flavor-mixing angle $\theta$ are shown in Fig.~\ref{f3(2F)}.
The experimental range ($200\pm20$ MeV)~\cite{Bugg:2004rj} is indicated by a blue band, overlayed with our $^3P_0$ predictions for direct comparison.
In the $f_3(2F)$ case, the total width follows a bell-shaped dependence on $\theta$: it is closest to the experimental central value near $\theta\simeq0$ and decreases as $|\theta|\to90^\circ$.
The prediction overlaps the RVUE band~\cite{Bugg:2004rj} for approximately $-28^\circ\lesssim\theta\lesssim22^\circ$.
At $\theta=0$, the predicted total width of $f_3(2F)$ lies within the experimental range. Hence, $\theta=0$ is again adopted for quantitative discussion. Opposite to $f_3(2F)$, $f_3^\prime(2F)$ reaches a minimum near $\theta=0$ and increases progressively as $|\theta|\to90^\circ$.
{For $\theta$ values ranging from $-28^\circ$ to $22^\circ$, the total decay width of $f_3^\prime(2F)$ remains between 147 MeV and 180 MeV. }

At $\theta=0$, the total width of $f_3(2F)$ is $210$~MeV. Most of the branching ratio is carried by $\pi a_2$, contributing 20\% of the width and a partial width of 40 MeV. $\pi a_4(2040)$ accounts for 13\% and 27 MeV, while $\pi \pi_2(1880)$ adds 10\% and 21 MeV. Table~\ref{1F2F} summarizes all other decay modes.

The calculated total width for $f_3^\prime(2F)$ amounts to 150 MeV when we take $\theta=0$. Ordering channels by yield, $K K_2^{*}$ supplies 21\% of the branching ratio with a 31 MeV partial width,
$K^{*}K^{*}(1410)$ supplies 12\% and 18 MeV, and $K K^{*}$ supplies 12\% and 17 MeV. The complete set of other channels is given in Table~\ref{1F2Fp}.

\begin{table}[htbp]
\renewcommand{\arraystretch}{1.4}
\centering
\caption{Total and partial decay widths of $f_3(1F)$ and $f_3(2F)$. ``Br''  denotes the branching ratio of the decay channel. The unit of width is MeV. Channels with branching ratios below 0.5\% are omitted.\label{1F2F}}

\[\begin{array}{cccccc}
\hline
\hline
 \text{} & {f_3(1F)} & \text{} & \text{} & {f_3(2F)} & \text{} \\
 \midrule[0.7pt]
 \text{Channel} & \text{Br($\%$)} & \text{Width} & \text{Channel} & \text{Br($\%$)} & \text{Width} \\
  \midrule[0.7pt]
 \text{Total} & 100 & 232 & \text{Total} & 100 & 207 \\
 \pi a_2 & 30.3 & 70.4 & \pi a_2 & 19.5 & 40.2 \\
 \rho \rho  & 16.9 & 39.2 & \pi a_4\text{(2040)} & 13.2 & 27.3 \\
 \rho b_1 & 11.2 & 25.9 & \pi \pi _2\text{(1880)} & 10 & 20.7 \\
 \omega h_1 & 9.48 & 22 & \pi a_2\text{(1700)} & 9.54 & 19.7 \\
 \pi a_1 & 8.1 & 18.8 & \pi a_1\text{(1640)} & 9.42 & 19.5 \\
 \omega \omega  & 8 & 18.6 & \rho b_1 & 8.57 & 17.7 \\
 \eta f_2 & 6.93 & 16.1 & \rho \rho  & 4.49 & 9.28 \\
 KK_2{}^* & 3.1 & 7.2 & \omega h_1 & 4.22 & 8.73 \\
 \pi \pi _2 & 2.49 & 5.78 & \eta f_2\text{(1640)} & 4 & 8.27 \\
 \pi a_0 & 0.924 & 2.14 & \eta f_2 & 2.78 & 5.74 \\
 KK^* & 0.888 & 2.06 & \text{$\rho \rho $(1450)} & 2.54 & 5.24 \\
 \pi a_2\text{(1700)} & 0.797 & 1.85 & \omega \omega  & 2.23 & 4.6 \\
 \text{...} & \text{...} & \text{...} & \pi a_0\text{(1740)} & 2.19 & 4.52 \\
 \text{...} & \text{...} & \text{...} & \text{$\omega \omega $(1420)} & 1.49 & 3.09 \\
 \text{...} & \text{...} & \text{...} & \pi \pi _2 & 1.35 & 2.8 \\
 \text{...} & \text{...} & \text{...} & \pi a_3\text{(2030)} & 0.697 & 1.44 \\
 \text{...} & \text{...} & \text{...} & \pi a_2\text{(2030)} & 0.65 & 1.34 \\
 \text{...} & \text{...} & \text{...} & KK^*\text{(1410)} & 0.64 & 1.32 \\
 \text{...} & \text{...} & \text{...} & \eta f_0\text{(1500)} & 0.556 & 1.15 \\
 \hline
 \hline
\end{array}\]
\end{table}

\begin{table}[htbp]
\renewcommand{\arraystretch}{1.4}
\centering
\caption{Total and partial decay widths of $f_3^\prime(1F)$ and $f_3^\prime(2F)$. ``Br''  denotes the branching ratio of the decay channel. The unit of width is MeV. Channels with branching ratios below 0.5\% are omitted.\label{1F2Fp}}
\[\begin{array}{cccccc}
\hline
\hline
 \text{} & {f_3^\prime(1F)} & \text{} & \text{} & {f_3^\prime(2F)} & \text{} \\
 \midrule[0.7pt]
 \text{Channel} &  \text{Br($\%$)}&  \text{Width}& \text{Channel} & \text{Br($\%$)}& \text{Width}  \\
 \midrule[0.7pt]
  \text{Total} & 99.7 & 111 & \text{Total} & 99.7 & 145 \\
 KK_2{}^* & 46.2 & 51.4 & KK_2{}^* & 21.4 & 31.2 \\
 KK^* & 22.9 & 25.5 & K^*K^*\text{(1410)} & 12.2 & 17.7 \\
 K^*K^* & 15.3 & 17.1 & KK^* & 11.5 & 16.7 \\
 \eta f'_2\text{(1525)} & 5.78 & 6.43 & KK_3{}^*\text{(1780)} & 9.61 & 14 \\
 KK_1^\prime & 4.08 & 4.54 & KK^*\text{(1410)} & 7.28 & 10.6 \\
 KK^*\text{(1410)} & 3.43 & 3.82 & K^*K_1 & 6.68 & 9.73 \\
 \phi \phi  & 1.25 & 1.4 & KK_1'\text{(2P)} & 4.93 & 7.17 \\
 \text{...} & \text{...} & \text{...} & K^*\text{K(1460)} & 4.84 & 7.04 \\
 \text{...} & \text{...} & \text{...} & K^*K_2{}^* & 4.48 & 6.52 \\
 \text{...} & \text{...} & \text{...} & K^*K^* & 4.48 & 6.51 \\
 \text{...} & \text{...} & \text{...} & KK_2{}^*\text{(1980)} & 3.13 & 4.55 \\
 \text{...} & \text{...} & \text{...} & KK_2\text{(1820)} & 2.65 & 3.86 \\
 \text{...} & \text{...} & \text{...} & K^*K_1^\prime & 1.04 & 1.51 \\
 \text{...} & \text{...} & \text{...} & \eta f'_2\text{(1525)} & 0.99 & 1.44 \\
 \text{...} & \text{...} & \text{...} & KK_1\text{(1790)} & 0.89 & 1.29 \\
 \text{...} & \text{...} & \text{...} & KK^*\text{(1680)} & 0.856 & 1.25 \\
 \text{...} & \text{...} & \text{...} & KK_1 & 0.747 & 1.09 \\
 \text{...} & \text{...} & \text{...} & K^*K_0{}^*\text{(1430)} & 0.72 & 1.05 \\
 \text{...} & \text{...} & \text{...} & \phi \phi  & 0.631 & 0.919 \\
\hline
\hline
\end{array}\]
 \end{table}

\begin{figure}[htbp]
\centering%
\includegraphics[scale=1]{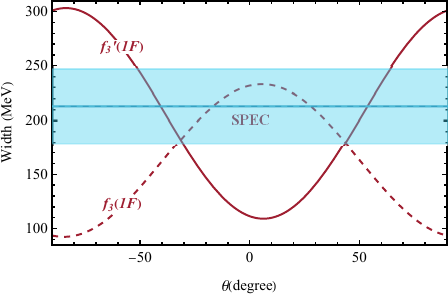}
\caption{The $\theta$ dependence of the total decay widths for $f_3(1F)$ and $f_3^\prime(1F)$, with the corresponding experimental data \cite{Anisovich:2000ut} (represented by a blue band) being presented alongside our theoretical calculation for comparison.}
\label{f3(1F)}
\end{figure}

\begin{figure}[htbp]
\centering%
\includegraphics[scale=1]{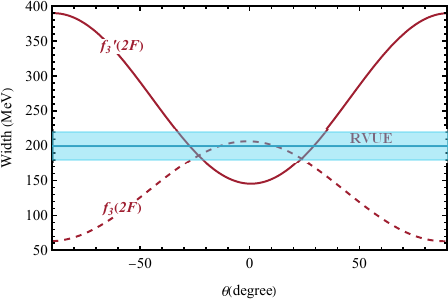}
\caption{The $\theta$ dependence of the total decay widths for $f_3(2F)$ and $f_3^\prime(2F)$, with the corresponding experimental data \cite{Bugg:2004rj} (represented by a blue band) being presented alongside our theoretical calculation for comparison.}
\label{f3(2F)}
\end{figure}


\subsubsection{Predictions of the second and the third  excited states of the light $3^{++}$ meson family}

We predict a total width of 110 MeV for $a_3(3F)$. The specific decay channels and their associated information is given in Table~\ref{3F4F}.
The four leading modes are $\pi\rho_3(2250)$ with a branching ratio of 25\% and a partial width of 28 MeV, $\pi\rho_3(1990)$ with 11\% and 12 MeV, $\pi f_2(1640)$ with 5.9\% and 6.5 MeV, and $\pi\rho_3(1690)$ with 4.8\% and 5.3 MeV, respectively.

For $a_3(4F)$ we obtain a total width of 79 MeV. The main decay modes are $\pi\rho_3(2250)$ with 10\% and 8.1 MeV, then $\pi\rho_3(1990)$ with 9.1\% and 7.2 MeV,
followed by $\pi f_2(1640)$ with 6.2\% and 4.9 MeV, and $\pi f_2(4P)$ with 4.1\% and 3.2 MeV. Remaining channels and details are given in Table~\ref{3F4F}.

Fig.~\ref{f3(3F)} displays the total widths of $f_3(3F)$, $f_3^\prime(3F)$, $f_3(4F)$, and $f_3^\prime(4F)$ as functions of the flavor-mixing angle $\theta$. $f_3(3F)$ and $f_3(4F)$ curves show a broad maximum near $\theta\simeq0$ and then decrease monotonically as $|\theta|\to90^\circ$, whereas $f_3'(3F)$ and $f_3'(4F)$ display an opposite trend, reaching a clear minimum around $\theta\simeq0$ and increasing toward the edges, peaking as $|\theta|\to90^\circ$.

Turning to $f_3(3F)$, the calculated total decay width is 97 MeV when we take $\theta=0$.
The main decay modes are $\pi a_2(1700)$ with a branching ratio of 20\% and a partial width of 19 MeV, $\pi a_2$ with 11\% and 10 MeV, $\pi a_4(2255)$ with 10\% and 10 MeV, and $\pi a_2(1950)$ with 10\% and 9.8 MeV. $\rho\rho$, $\omega\omega$, $\pi a_1(3P)$, and $\rho b_1$ remain important contributors to the $f_3(3F)$ width. The rest of the channels, together with specifics, are listed in Table~\ref{2H4F}.

In the hidden-strangeness sector, $f_3^\prime(3F)$ has a total width of 130 MeV at $\theta=0$.
Table~\ref{2H4F} lists the specific decay channels and the related decay information, among which $K K_2^{*}$ contributes 14\% with a partial width of 17 MeV, $K K_3^{*}(1780)$ contributes 9.3\% with 12 MeV, $K K_2^{*}(1980)$ contributes 8.5\% with 11 MeV, and $K K^{*}$ contributes 7.6\% with 9.7 MeV. Besides, $f_3^\prime(3F)$ can also decay into $K K^{*}(1410)$, $K^{*}K_2^{*}$, $K^{*}K_1$, and $K^{*}K^{*}(1410)$.

As for $f_3(4F)$, the predicted width is 75 MeV when taking $\theta=0$.
The main decay modes are $\pi a_2(1700)$ with a branching ratio of 14\% and a partial width of 11 MeV, $\pi a_2(1950)$ with 10\% and 7.7 MeV, $\pi a_2(2175)$ with 8.1\% and 6.1 MeV, and $\pi a_4(2255)$ with 7.1\% and 5.3 MeV. Other notable contributions arise from $\rho\rho$, $\pi a_2$, $\rho\rho(1450)$, and $\pi\pi_2(1880)$. See Table~\ref{2H4F} for other channels.

Finally, for $f_3^\prime(4F)$ we obtain a total width of 130 MeV at $\theta=0$. The four main strange final states are $K K_2^{*}$ with a branching ratio of 8.3\% and a partial width of 11 MeV, $K K_3^{*}(1780)$ with 6.8\% and 8.8 MeV, $K K_2^{*}(3P)$ with 6.2\% and 7.9 MeV, and $K^{*}K_2^{*}$ with 6.0\% and 7.8 MeV. Further sizable contributions come from $K K_2^{*}(1980)$, $K K^{*}$, $K K^{*}(1410)$, and $K K_3^{*}(2D)$, as tabulated in Table~\ref{2H4F}.

\begin{table}[htbp]
\renewcommand{\arraystretch}{1.2}
\vspace{-0.8cm}
\centering
\caption{Total and partial decay widths of $a_3{(3F)}$ and $a_3(4F)$. The unit of width is MeV. Channels with branching ratios below 0.5\% are omitted.\label{3F4F}}
\[\begin{array}{cccccc}
\hline
\hline
 \text{} & {a_3(3F)} & \text{} & \text{} & {a_3(4F)} & \text{} \\
 \midrule[0.7pt]
 \text{Channel}& \text{Br(\%)} & \text{Width}  & \text{Channel} & \text{Br(\%)} & \text{Width} \\
 \midrule[0.7pt]
 \text{Total} & 100 & 111 & \text{Total} & 100 & 78.8 \\
 \pi \rho _3\text{(2250)} & 24.8 & 27.5 & \pi \rho _3\text{(2250)} & 10.3 & 8.09 \\
 \pi \rho _3\text{(1990)} & 11 & 12.2 & \pi \rho _3\text{(1990)} & 9.13 & 7.19 \\
 \pi f_2\text{(1640)} & 5.85 & 6.48 & \pi f_2\text{(1640)} & 6.2 & 4.88 \\
 \pi \rho _3\text{(1690)} & 4.8 & 5.31 & \pi f_2\text{(4P)} & 4.08 & 3.22 \\
 \pi \rho  & 4.39 & 4.86 & \pi \rho  & 3.89 & 3.06 \\
 \rho \omega  & 4.09 & 4.53 & \rho \omega  & 3.14 & 2.47 \\
 \pi f_2 & 3.81 & 4.22 & \pi f_2\text{(1950)} & 3.13 & 2.47 \\
 \text{$\pi \rho $(1450)} & 3.13 & 3.46 & \text{$\pi \rho $(1450)} & 2.88 & 2.27 \\
 \pi f_2\text{(1950)} & 2.89 & 3.2 & \text{$\pi \rho $(1900)} & 2.8 & 2.21 \\
 \text{$\pi \rho $(1900)} & 2.87 & 3.18 & \rho a_2 & 2.57 & 2.02 \\
 \rho a_2 & 2.69 & 2.97 & \pi f_2 & 2.21 & 1.74 \\
 \rho a_1 & 2.18 & 2.41 & \pi \rho _3\text{(1690)} & 2.13 & 1.67 \\
 \pi \rho _2\text{(2D)} & 1.99 & 2.2 & \text{$\rho \omega $(1420)} & 2.07 & 1.63 \\
 \text{$\rho \pi $(1300)} & 1.91 & 2.11 & \text{$\pi \rho $(4S)} & 2.02 & 1.59 \\
 \rho h_1 & 1.73 & 1.92 & \text{$\rho \pi $(1300)} & 1.99 & 1.57 \\
 \pi b_1\text{(1960)} & 1.55 & 1.72 & \pi \rho _2\text{(2D)} & 1.86 & 1.47 \\
 \pi \rho _2\text{(1D)} & 1.43 & 1.58 & \pi \rho _2\text{(3D)} & 1.75 & 1.38 \\
 \pi f_3\text{(2050)} & 1.32 & 1.46 & \rho \pi _2 & 1.62 & 1.27 \\
 \pi f_0\text{(1500)} & 1.24 & 1.37 & \rho a_1 & 1.58 & 1.24 \\
 \pi f_1\text{(3P)} & 1.11 & 1.23 & \text{$\omega \rho $(1450)} & 1.45 & 1.14 \\
 \omega b_1 & 0.893 & 0.989 & \pi \rho _2\text{(1D)} & 1.36 & 1.07 \\
 \pi f_1 & 0.851 & 0.942 & \rho h_1 & 1.33 & 1.05 \\
 \pi \eta _2 & 0.82 & 0.908 & \pi b_1\text{(1960)} & 1.32 & 1.04 \\
 \text{$\rho \omega $(1420)} & 0.798 & 0.884 & \eta a_2\text{(1950)} & 1.24 & 0.974 \\
 \pi f_4\text{(2300)} & 0.738 & 0.817 & \pi f_3\text{(2300)} & 1.2 & 0.948 \\
 \text{$\pi \rho $(1700)} & 0.677 & 0.75 & \text{$\pi \rho $(2000)} & 1.16 & 0.911 \\
 \text{$\pi \rho $(4S)} & 0.632 & 0.699 & \rho h_1\text{(1595)} & 1.14 & 0.898 \\
 \text{$\pi \rho $(2000)} & 0.625 & 0.692 & \rho a_1\text{(1640)} & 1.12 & 0.884 \\
 \pi f_4\text{(2050)} & 0.608 & 0.673 & \pi \rho _3\text{(2G)} & 1.05 & 0.829 \\
 \pi \rho _3\text{(1G)} & 0.573 & 0.634 & \pi f_1 & 1.05 & 0.826 \\
 KK_3{}^*\text{(1780)} & 0.539 & 0.596 & \pi b_3\text{(2244)} & 1.03 & 0.811 \\
 \eta a_1 & 0.536 & 0.593 & \rho a_2\text{(1700)} & 1.01 & 0.795 \\
 \pi b_3\text{(2244)} & 0.513 & 0.568 & \pi f_4\text{(3F)} & 0.965 & 0.76 \\
 \eta a_2\text{(1700)} & 0.505 & 0.558 & \pi f_4\text{(2300)} & 0.883 & 0.696 \\
 \text{...} & \text{...} & \text{...} & \pi \eta _2 & 0.794 & 0.626 \\
 \text{...} & \text{...} & \text{...} & \omega b_1 & 0.762 & 0.6 \\
 \text{...} & \text{...} & \text{...} & \pi b_1\text{(2240)} & 0.757 & 0.596 \\
 \text{...} & \text{...} & \text{...} & \pi f_3\text{(2050)} & 0.703 & 0.554 \\
 \text{...} & \text{...} & \text{...} & \pi f_1\text{(4P)} & 0.701 & 0.552 \\
 \text{...} & \text{...} & \text{...} & \rho \omega _2\text{(1D)} & 0.63 & 0.496 \\
 \text{...} & \text{...} & \text{...} & \omega \rho _2\text{(1D)} & 0.613 & 0.483 \\
 \text{...} & \text{...} & \text{...} & \pi \eta _2\text{(2D)} & 0.602 & 0.474 \\
 \text{...} & \text{...} & \text{...} & \text{$\pi \rho $(3D)} & 0.536 & 0.422 \\
 \text{...} & \text{...} & \text{...} & \pi b_1 & 0.534 & 0.42 \\
 \text{...} & \text{...} & \text{...} & \eta a_1\text{(1640)} & 0.522 & 0.411 \\
 \text{...} & \text{...} & \text{...} & \eta a_1 & 0.519 & 0.409 \\
 \text{...} & \text{...} & \text{...} & \rho \omega _3\text{(1670)} & 0.519 & 0.408 \\
 \text{...} & \text{...} & \text{...} & \pi f_1\text{(2P)} & 0.512 & 0.403 \\
\hline
\hline
\end{array}\]
\end{table}

\begin{figure}[htbp]
\centering%
\includegraphics[scale=1]{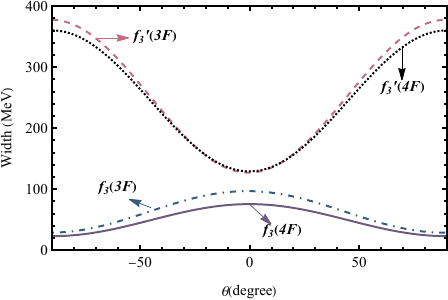}
\caption{The $\theta$ dependence of the total decay widths for $f_3(3F)$, $f_3^\prime(3F)$, $f_3(4F)$, and $f_3^\prime(4F)$.}
\label{f3(3F)}
\end{figure}

\begin{table*}[htbp]
\renewcommand{\arraystretch}{1.2}
\vspace{-0.8cm}
\centering
\caption{Total and partial decay widths of $f_3{(3F)}$, $f_3^\prime{(3F)}$, $f_3{(4F)}$, and $f_3^\prime(4F)$. The unit of width is MeV. Channels with branching ratios below 0.5\% are omitted.\label{2H4F}}
\[\begin{array}{ccccccccccccccccc}
\hline
\hline
 \text{} & {f_3(3F)} & \text{} & \text{} & {f_3^\prime(3F)} & \text{}& \text{} & {f_3(4F)} & \text{} & \text{} & {f_3^\prime(4F)} & \text{} \\
 \midrule[0.7pt]
 \text{Channel} & \text{Br($\%$)} & \text{Width} & \text{Channel} & \text{Br($\%$)} & \text{Width} & \text{Channel} & \text{Br($\%$)} & \text{Width}
& \text{Channel} & \text{Br($\%$)} & \text{Width} \\
\hline
 \text{Total} & 100 & 96.7 & \text{Total} & 99.9 & 127 & \text{Total} & 99.9 & 75.4 & \text{Total} & 99.9 & 129 \\
 \pi a_2\text{(1700)} & 19.5 & 18.8 & KK_2{}^* & 13.7 & 17.4 & \pi a_2\text{(1700)} & 14 & 10.6 & KK_2{}^* & 8.26 & 10.7 \\
 \pi a_2 & 10.5 & 10.1 & KK_3{}^*\text{(1780)} & 9.25 & 11.8 & \pi a_2\text{(1950)} & 10.2 & 7.71 & KK_3{}^*\text{(1780)} & 6.78 & 8.75 \\
 \pi a_4\text{(2255)} & 10.4 & 10 & KK_2{}^*\text{(1980)} & 8.46 & 10.8 & \pi a_2\text{(2175)} & 8.05 & 6.07 & KK_2{}^*\text{(3P)} & 6.15 & 7.94
\\
 \pi a_2\text{(1950)} & 10.1 & 9.81 & KK^* & 7.6 & 9.66 & \pi a_4\text{(2255)} & 7.08 & 5.34 & K^*K_2{}^* & 6.04 & 7.8 \\
 \rho \rho  & 9.57 & 9.26 & KK^*\text{(1410)} & 7.6 & 9.66 & \rho \rho  & 6.76 & 5.1 & KK_2{}^*\text{(1980)} & 5.87 & 7.58 \\
 \omega \omega  & 4.58 & 4.43 & K^*K_2{}^* & 5.84 & 7.43 & \pi a_2 & 5.3 & 4 & KK^* & 5.39 & 6.96 \\
 \pi a_1\text{(3P)} & 3.82 & 3.69 & K^*K_1 & 5.71 & 7.26 & \text{$\rho \rho $(1450)} & 4.78 & 3.61 & KK^*\text{(1410)} & 4.89 & 6.32 \\
 \pi a_1 & 3.48 & 3.36 & K^*K^*\text{(1410)} & 5.58 & 7.09 & \pi a_1 & 4.43 & 3.34 & KK_3{}^*\text{(2D)} & 4.58 & 5.91 \\
 \rho b_1 & 3.27 & 3.16 & K^*K^* & 4.48 & 5.69 & \omega \omega  & 3.18 & 2.4 & K^*K^* & 4.46 & 5.75 \\
 \pi a_4\text{(2040)} & 3.21 & 3.11 & KK^*\text{(3S)} & 4.39 & 5.58 & \pi a_4\text{(3F)} & 3.02 & 2.28 & K^*K_1 & 4.1 & 5.29 \\
 \eta f_2\text{(1640)} & 2.58 & 2.49 & K^*\text{K(1460)} & 4.29 & 5.45 & \rho b_1 & 2.5 & 1.89 & KK^*\text{(3S)} & 3.23 & 4.16 \\
 \pi \pi _2 & 2.24 & 2.17 & KK_2'\text{(2D)} & 3.63 & 4.62 & \pi \pi _2\text{(1880)} & 2.43 & 1.84 & K^*K^*\text{(1410)} & 3.03 & 3.91 \\
 \omega h_1 & 1.87 & 1.81 & K^*K_1^\prime & 2.83 & 3.6 & \pi a_1\text{(4P)} & 2.2 & 1.66 & K^*K_1^\prime & 2.98 & 3.85 \\
 \text{$\rho \rho $(1450)} & 1.82 & 1.76 & KK_2\text{(1820)} & 2.25 & 2.86 & \text{$\omega \omega $(1420)} & 2.08 & 1.57 & K^*\text{K(1460)} & 2.87
& 3.7 \\
 \pi \pi _2\text{(1880)} & 1.71 & 1.65 & KK_3{}^*\text{(2D)} & 1.84 & 2.34 & \rho \rho _2\text{(1D)} & 1.97 & 1.49 & KK_4{}^*\text{(2045)} & 2.68
& 3.47 \\
 \pi a_2\text{(2175)} & 1.23 & 1.19 & KK_1'\text{(3P)} & 1.47 & 1.87 & \pi \pi _2 & 1.84 & 1.39 & K_1K_2{}^* & 2.56 & 3.31 \\
 \pi a_2\text{(2F)} & 1.13 & 1.1 & KK_4{}^*\text{(2045)} & 1.37 & 1.74 & \eta f_2\text{(1640)} & 1.62 & 1.22 & KK^*\text{(4S)} & 2.41 & 3.11 \\
 \rho \rho _2\text{(1D)} & 1.11 & 1.07 & KK_1^\prime & 1.26 & 1.6 & \rho b_1\text{(2P)} & 1.54 & 1.16 & KK_2'\text{(2D)} & 2.22 & 2.87 \\
 \text{$\omega \omega $(1420)} & 0.864 & 0.836 & K^*K_1'\text{(2P)} & 1.15 & 1.46 & \eta f_2\text{(1950)} & 1.46 & 1.1 & KK_2\text{(1820)} & 1.92
& 2.48 \\
 \pi a_0\text{(3P)} & 0.854 & 0.826 & K^*K_0{}^*\text{(1430)} & 0.857 & 1.09 & \omega h_1 & 1.32 & 0.997 & KK_1^\prime & 1.74 & 2.25 \\
 \eta f_2 & 0.809 & 0.782 & \phi \phi  & 0.802 & 1.02 & \rho \rho _3\text{(1690)} & 1.28 & 0.967 & K^*K_2\text{(1820)} & 1.7 & 2.2 \\
 KK_3{}^*\text{(1780)} & 0.617 & 0.596 & KK_1\text{(1790)} & 0.796 & 1.01 & \pi a_4\text{(2040)} & 1.17 & 0.886 & K^*\text{K(1830)} & 1.66 & 2.14
\\
 \pi a_1\text{(1640)} & 0.572 & 0.553 & KK^*\text{(1680)} & 0.659 & 0.838 & \omega h_1\text{(1595)} & 1.08 & 0.812 & K^*K_1'\text{(2P)} & 1.13 &
1.46 \\
 \text{...} & \text{...} & \text{...} & KK_1{}^*\text{(2D)} & 0.594 & 0.755 & \pi a_2\text{(3F)} & 0.9 & 0.679 & K^*K_1\text{(1790)} & 1.09 & 1.4
\\
 \text{...} & \text{...} & \text{...} & \text{...} & \text{...} & \text{...} & \pi a_1\text{(1640)} & 0.641 & 0.483 & KK_4{}^*\text{(2F)} & 0.966
& 1.25 \\
 \text{...} & \text{...} & \text{...} & \text{...} & \text{...} & \text{...} & \omega \omega _2\text{(1D)} & 0.64 & 0.483 & K^*K_3{}^*\text{(1780)}
& 0.96 & 1.24 \\
 \text{...} & \text{...} & \text{...} & \text{...} & \text{...} & \text{...} & \pi a_3\text{(3F)} & 0.552 & 0.416 & K_1K^*\text{(1410)} & 0.717 &
0.926 \\
 \text{...} & \text{...} & \text{...} & \text{...} & \text{...} & \text{...} & \pi a_0\text{(3P)} & 0.518 & 0.391 & KK_2\text{(2250)} & 0.681 & 0.88
\\
 \text{...} & \text{...} & \text{...} & \text{...} & \text{...} & \text{...} & \omega \omega _3\text{(1670)} & 0.51 & 0.385 & K^*K_0{}^*\text{(1430)}
& 0.636 & 0.822 \\
 \text{...} & \text{...} & \text{...} & \text{...} & \text{...} & \text{...} & \text{...} & \text{...} & \text{...} & KK_1{}^*\text{(2D)} & 0.612
& 0.79 \\
 \text{...} & \text{...} & \text{...} & \text{...} & \text{...} & \text{...} & \text{...} & \text{...} & \text{...} & KK_3'\text{(1F)} & 0.583 &
0.753 \\
 \text{...} & \text{...} & \text{...} & \text{...} & \text{...} & \text{...} & \text{...} & \text{...} & \text{...} & K^*K_2\text{(1770)} & 0.52
& 0.672 \\
 \hline
 \hline
\end{array}\]
\end{table*}

\section{Conclusion}\label{sec4}
In this work, we have systematically studied the mass spectrum and OZI-allowed two-body strong decays of the $3^{++}$ family, and proposed the assignments for the states $a_3(1875)$, $a_3(2030)$, $a_3(2275)$, $f_3(2050)$, and $f_3(2300)$.
\par

Our analysis indicates that $a_3(1875)$ and $a_3(2030)$ likely correspond to the same state, namely the $a_3$ ground state, while $f_3(2050)$ is the ground state of the $f_3$ family. Likewise, $a_3(2275)$ and $f_3(2300)$ can be interpreted as the first radial excitation of the $3^{++}$ meson family, with our predicted masses and widths being consistent with existing measurements.
\par

In addition, we predict the following masses and total widths:
\begin{equation*}
\begin{cases}
M(a_3(3F))= 2454~\text{MeV}\\
\Gamma(a_3(3F))= 111~\text{MeV}
\end{cases}
\begin{cases}
M(a_3(4F))= 2631~\text{MeV}\\
\Gamma(a_3(4F))= 78.8~\text{MeV}
\end{cases},
\end{equation*}
\begin{equation*}
\begin{cases}
M(f_3(3F))= 2454~\text{MeV}\\
\Gamma(f_3(3F))= 96.7~\text{MeV}
\end{cases}
\begin{cases}
M(f_3(4F))= 2631~\text{MeV}\\
\Gamma(f_3(4F))= 75.4~\text{MeV}
\end{cases},
\end{equation*}
\begin{equation*}
\begin{cases}
M(f_3^\prime(1F))= 2166~\text{MeV}\\
\Gamma(f_3^\prime(1F))= 111~\text{MeV}
\end{cases}
\begin{cases}
M(f_3^\prime(2F))= 2452~\text{MeV}\\
\Gamma(f_3^\prime(2F))= 145~\text{MeV}
\end{cases},
\end{equation*}
\begin{equation*}
\begin{cases}
M(f_3^\prime(3F))= 2682~\text{MeV}\\
\Gamma(f_3^\prime(3F))= 127~\text{MeV}
\end{cases}
\begin{cases}
M(f_3^\prime(4F))= 2870~\text{MeV}\\
\Gamma(f_3^\prime(4F))= 129~\text{MeV}
\end{cases}.
\end{equation*}

{We also provide their main decay modes.
The $a_3$ family has the following leading modes: at $1F$ it decays primarily to $\pi \rho_3(1690)$, at $2F$ the strength shifts to $\pi \rho_3(1990)$, whereas the $3F$ and $4F$ states favor $\pi \rho_3(2250)$. $f_3$ is led by $\pi a_2$ at $1F$ and $2F$, with the preference moving to $\pi a_2(1700)$ for $3F$ and $4F$. The $f_3^\prime$--both the ground state and higher excitations--are characterized by $KK_2^*$ as the principal final state. These predictions may serve as valuable guidance for experimental identification and future searches of the $a_3$, $f_3$, and $f_3^\prime$ families.}

We anticipate that forthcoming experimental measurements will play a decisive role in clarifying the nature of the observed light mesons with $I^GJ^{PC}=1^{-}3^{++}$ and $I^GJ^{PC}=0^{+}3^{++}$, and in testing the predictions presented here.

\begin{acknowledgments}
This work is supported by the National Natural Science Foundation of China under Grants   No.~11965016, and No.~12247101, and by the Natural Science Foundation of Qinghai Province under Grant No. 2022-ZJ-939Q, the Fundamental Research Funds for the Central Universities (Grant No. lzujbky-2024-jdzx06).
\end{acknowledgments}
\bibliographystyle{apsrev4-1}
\bibliography{hepref}
\end{document}